\documentclass[preprint]{aastex}
\usepackage{graphicx}
\usepackage{subfigure}
\usepackage{mathrsfs,amssymb}
\usepackage{amsmath}
\usepackage{natbib}
\usepackage{color}
\usepackage{ulem}
\usepackage{bm}

\def\gtsima{$\; \buildrel > \over \sim \;$}
\def\ltsima{$\; \buildrel < \over \sim \;$}
\def\gsim{\lower.5ex\hbox{\gtsima}}
\def\lsim{\lower.5ex\hbox{\ltsima}}
\def\simgt{\lower.5ex\hbox{\gtsima}}
\def\simlt{\lower.5ex\hbox{\ltsima}}
\def\simpr{\lower.5ex\hbox{\prosima}}
						
\begin{document}

\title{The contribution of high redshift galaxies to the Near-Infrared Background} 

\author{Bin Yue\altaffilmark{1,3,5},
Andrea Ferrara\altaffilmark{1,6},
Ruben Salvaterra\altaffilmark{2},
Xuelei Chen\altaffilmark{3,4}
}
\altaffiltext{1}{Scuola Normale Superiore, Piazza dei Cavalieri 7, I-56126 Pisa, Italy}
\altaffiltext{2}{INAF, IASF Milano, via E. Bassini 15, I-20133 Milano, Italy}
\altaffiltext{3}{National Astronomical Observatories, Chinese Academy of Sciences,
20A Datun Road, Chaoyang, Beijing 100012, China}
\altaffiltext{4}{Center of High Energy Physics, Peking University, Beijing 100871, China}
\altaffiltext{5}{Graduate University of Chinese Academy of Sciences, Beijing 100049,China}
\altaffiltext{6}{Centennial B. Tinsley Professor, University of Texas, Austin, USA}
						
\begin{abstract}
Several independent measurements have confirmed the existence of fluctuations 
($\delta F_{\rm obs}\approx 0.1~\rm nW/m^{2}/sr$ at $3.6~\rm \mu m$) up to degree 
angular scales in the source-subtracted Near InfraRed Background (NIRB) 
whose origin is unknown. By combining high resolution cosmological N-body/hydrodynamical
simulations with an analytical 
model, and by matching galaxy Luminosity Functions (LFs) and the constraints 
on reionization simultaneously, we predict the NIRB 
absolute flux and fluctuation amplitude produced by high-$z$ ($z > 5$) galaxies 
(some of which harboring Pop III stars, shown to 
provide a negligible contribution). This strategy also allows 
us to make an empirical determination of 
the evolution of ionizing photon escape fraction: we find $f_{\rm esc} = 1$ at $z \ge 11$, decreasing to $\approx 0.05$ at $z = 5$.  In the wavelength range $1.0-4.5~\rm \mu m$, the predicted cumulative flux is  $F =0.2-0.04~\rm nW/m^2/sr$. 
However, we find that the radiation from high-$z$ galaxies (including those undetected by current surveys) is insufficient to 
explain the amplitude of the observed fluctuations: at $l=2000$, the fluctuation level due to $z > 5$ galaxies is $\delta F = 0.01-0.002~\rm nW/m^2/sr$, with a relative wavelength-independent 
amplitude $\delta F/F = 4$\%. The source of the missing power remains unknown. This might indicate that an
unknown component/foreground, with a clustering signal very similar to that of high-$z$ galaxies,
dominates the source-subtracted NIRB fluctuation signal. 
\end{abstract}

\maketitle


\section {INTRODUCTION}\label{intro}
  
Observations of high redshift galaxies are essential to understand cosmic reionization. Although current surveys have reached redshifts $\sim8-10$ \citep{2010ApJ...709L.133B,2011Natur.469..504B,
2011ApJ...737...90B}, it is generally believed that the sources
detected so far, usually rare and bright galaxies, are not the dominant contributors to reionization
\citep{2007MNRAS.380L...6C,2011MNRAS.414.1455L,2012MNRAS.420.1606J,2012ApJ...758...93F}. 
Instead, reionization is likely
powered by the large number of galaxies that are still below the detection limit. 

Even without detecting these faint galaxies individually, their cumulative radiations
may still tell us much about their
properties. Indeed, the bulk of their emission, mostly in the band between the Lyman limit and the
visible light, is redshifted into the near InfraRed (IR)
at present time. Therefore, the Near InfraRed Background (NIRB) which is obtained 
after removing the contributions from the Solar system, the Milky Way and low-$z$ galaxies, could provide a wealth of information on high redshift galaxies, such as their integrated emissivity and large scale clustering properties. 

The NIRB measurement has a history dating back to 
more than two decades (see the review by \citealt{2005PhR...409..361K}).
Early measurements gave a NIRB flux $\simgt 10~ \rm nW/m^2/sr$ \citep{1998ApJ...508L...9D,2000ApJ...536..550G,
2000LNP...548...96M,2001ApJ...555..563C,2005ApJ...626...31M}.
These works showed that a non-zero residual remains after the foreground 
and the emission from known galaxies are removed \citep{2001ApJ...550L.137T,2005ApJ...626...31M}.
\cite{2003MNRAS.339..973S} and \cite{2002MNRAS.336.1082S} suggested 
that Pop III stars could possibly be the sources of such leftover signal. If true, this residual 
would be an exquisite tool to study Pop III stars. However, not long afterwards,
\cite{2005MNRAS.359L..37M} and \cite{2006MNRAS.367L..11S} found that this scenario 
needs a very high star formation efficiency and may overpredict
the high-$z$ dropouts galaxies\footnote{A significant contribution
from high-$z$ mini-quasars powered by accretion on to intermediate
mass black holes with spectra similar to local Ultra-Luminous X-ray sources is disfavored 
on the basis of the unresolved X-ray background intensity constraints \citep{2005MNRAS.362L..50S},
unless these objects are highly absorbed by the surrounding gas.}.
To solve this problem, we need either an alternative theoretical explanation
(proved hard to be found), or a more accurate determination 
of the residual flux, or both.

Due to the difficulties in foreground subtraction \citep{2005ApJ...635..784D},
in recent observational works more attentions are
paid to the angular fluctuations.
In such observations,
the influence of strong but smooth 
foregrounds, such as the zodiacal light,
is reduced, and one can also infer
the large scale clustering properties of the 
unknown sources \citep{2002ApJ...579L..53K,2004ApJ...608....1K,2003MNRAS.342L..25M,
2004ApJ...606..611C,2005ApJ...626...31M,2006MNRAS.368L...6S}.
The recent measurements \citep{2005Natur.438...45K,2007ApJ...654L...5K,
2012ApJ...753...63K,2005ApJ...626...31M,2011ApJ...742..124M,2007ApJ...657..669T,
2007ApJ...666..658T, 2012Natur.490..514C}
have obtained angular power spectra of the source-subtracted NIRB 
(i.e. all resolved galaxies have been removed) at wavelengths
from $1.1~\rm \mu m$ to $8~\rm \mu m$. 
These angular power spectrum measurements
show that 
the sources have a large clustering signal up to degree scales. 

The source-subtracted NIRB fluctuations are 
found to be much higher than the theoretically predicted contribution from low-$z$ faint galaxies.
Although \cite{2005Natur.438...45K, 2007ApJ...654L...1K,2011ApJ...742..124M,2012ApJ...753...63K}
favored a scenario in which the observed fluctuations come from Pop III stars, 
\cite{2012ApJ...756...92C} showed that, to be consistent with 
the electron scattering optical depth measured
by WMAP \citep{2011ApJS..192...18K},
the contribution from 
high-$z$ galaxies (including Pop III stars) must be smaller by
at least an order of magnitude than what is observed. 
Instead, \cite{2012Natur.490..514C} suggested recently 
that 
a large fraction of the observed NIRB fluctuations
comes from the diffuse light of intrahalo stars at
intermediate redshifts ($z~\sim$ 1 to 4).
While an intriguing idea, this explanation relies on the
poorly known fraction and spectral energy distribution of intrahalo stars. It also predicts, contrary
to the faint distant galaxies hypothesis, that fluctuations induced by the much closer intrahalo stars
should extend into the optical bands, where the light from first galaxies is blanketed by intervening
intergalactic neutral hydrogen.
Numerical simulations by \cite{2010ApJ...710.1089F,2012ApJ...750...20F} stressed 
the importance of nonlinear effects in theoretical calculations as a possible way to reconcile 
the theory with data. 

There have also been proposals that the sources of these fluctuations are lower 
redshift galaxies
\citep{2007ApJ...666..658T,2007ApJ...659L..91C,
2008ApJ...681...53C},
but this possibility has become less attractive by now. Indeed, \cite{2012ApJ...752..113H} recently
reconstructed the emissivity history from the luminosity functions (LFs) of observed galaxies, 
and found that the fluctuations from the known galaxy population below
the detection limit are unable to account for the observed clustering signal on sub-degree angular scales.

To make further progress, it is essential to make more accurate predictions of 
the NIRB contributed by Pop III stars and the galaxies before reionization, 
using models which are consistent with {\it all} current observational constraints, 
including both the high redshift LFs and reionization. 

In this paper, we attempt to make the most detailed theoretical NIRB model developed
so far, with predictions
on both the absolute flux and the angular power spectrum contributed by high-$z$ galaxies.
To do this, we used a simulation with detailed treatment of the relevant physics of star/galaxy formation,
including gas dynamics, radiative cooling, supernova explosion, photoionization and heating,
and especially 
a detailed treatment of chemical feedback \citep{2007MNRAS.382.1050T,2007MNRAS.382..945T}.  
The LFs of high redshift galaxies in the simulation match remarkably well
with observations, this is the starting point of our NIRB model.  

The layout of the paper is as follows. In Section \ref{methods} we introduce the simulation, and describe the steps 
to calculate the NIRB absolute flux and the angular power spectrum. In Section 
\ref{results} we present our results and compare them with observations. 
Conclusions are presented in Section \ref{conclusions}.
In Appendix \ref{appendix} we compare the different 
approximate solutions for the analytical calculation of the emissivity.
Throughout this paper,
we use the same cosmological parameters as in
\cite{2011MNRAS.414..847S}: $\Omega_m$=0.26, $\Omega_\Lambda$=0.74,
$h$=0.73, $\Omega_b$=0.041, $n=1$ and $\sigma_8$=0.8.
The transfer function is from \cite{1998ApJ...496..605E}. Magnitudes
are given in the AB system.

\section{METHOD}\label{methods}
\subsection{The Absolute Flux}

At $z=0$, the cumulative flux 
of the NIRB observed 
at frequency $\nu_0$ is the integrated contribution of 
sources whose emission is shifted into a 
band of central frequency $\nu_0$. Following \cite{2006MNRAS.368L...6S}, we write it as
\begin{eqnarray}
F&=&\nu_0I_{\nu_0} \nonumber \\
&=&\nu_0\int_{z_{\rm min}}^{z_{\rm max}}\epsilon(\nu,z){\rm e}^{-\tau_{\rm eff}(\nu_0,z)}\frac{dr_p}{dz}dz \nonumber\\
&=&\int_{z_{\rm min}}^{z_{\rm max}}cdz\frac{\nu\epsilon(\nu,z){\rm e}^{-\tau_{\rm eff}(\nu_0,z)}}{H(z)(1+       z)^2},
\label{I0}
\end{eqnarray}
where $r_p$ is the proper distance,
$\nu=(1+z)\nu_0$ is the rest frame 
frequency, 
$\epsilon(\nu,z)$ is the comoving specific emissivity, 
$H(z)$ is the Hubble parameter given by $H(z)=H_0\sqrt{\Omega_m(1+z)^3+\Omega_\Lambda}$
in a flat $\Lambda$CDM cosmology,
$c$ is the speed of light. The effective optical depth of absorbers between 
redshift 0 and $z$, $\tau_{\rm eff}$, is composed of two parts: the line absorption
and the continuum absorption;
we use the expressions in \cite{2003MNRAS.339..973S}.

We calculate the emissivity (see also Appendix for further discussions on subtleties related to the
various approximations used in the literature) from the results of the simulation
presented in \cite{2011MNRAS.414..847S}, which includes a detailed
treatment of chemical enrichment developed by \cite{2007MNRAS.382.1050T}.
In our model, both Pop II stars and Pop III stars are assumed to follow the Salpeter
initial mass function (IMF) \citep{1955ApJ...121..161S}, for Pop II
stars the mass range is $0.1-100~M_\odot$, while for Pop III stars 
the mass range is set to be $100-500~M_\odot$.
Some recent works indicate that Pop III stars may not be 
so massive as was predicted previously,
but may be limited to $\lesssim50~M_\odot$ \citep{2011Sci...334.1250H}.
Our choice then corresponds to an upper limit to the contribution of these
sources.
Using this simulation, \cite{2011MNRAS.414..847S} generated the LFs
of galaxies down to the magnitude far below the current observation 
limits at high redshifts. In the redshift range $5 < z < 10$, 
the simulated LFs match the observed  
ones almost perfectly in the overlapping luminosity range.

Suppose the specific 
luminosity of the $i$-th galaxy in the simulation box is $L^i_\nu(z)$
at redshift $z$, the comoving specific emissivity is then
\footnote{The dust absorption in the 
host galaxy may lower its luminosity in the rest frame UV band,
therefore reduce their contribution to the NIRB. However, as shown 
by \cite{2011MNRAS.414..847S}, for high redshift
dwarf galaxies considered by us here, this effect is negligible.
The absorption is possible important for very massive objects in the 
bright end of LFs, but 
their contribution to the emissivity is very small.
The lack of significant dust absorption is further supported by
the observed very blue UV-continuum slopes of high-$z$ galaxies 
reported in
\cite{2009ApJ...705..936B}.
Therefore, we ignore the effect of dust in current calculations.
Furthermore, we will show latter that the 
high-$z$ galaxies are inadequate to 
explain the amplitude of the observed fluctuations,
considering the dust absorption would only strength this point.
}
\begin{equation}
\epsilon(\nu,z)=\frac{1}{4\pi}\frac{\sum_{i=1}^NL^i_\nu(z)}{V},
\label{e}
\end{equation}
where $V$ is the comoving volume of the simulation, $N$ is the total number of galaxies in 
the simulation box at redshift $z$. 

In the emissivity calculation, we must correct for rare 
bright galaxies that are not caught by the simulation due to the finite box size
($10~h^{-1}$Mpc). We follow the steps in \cite{2011MNRAS.414..847S}.
We first calculate the absolute magnitude corresponding to the mean luminosity of the two  
brightest galaxies in the simulation box, 
$M_{\rm UV,up}$. The contribution (to be added to the numerator in Eq. (\ref{e}))
from galaxies brighter than this magnitude 
is obtained by integration
\begin{equation}
L^{\rm corr}_\nu(z)=V\int_{-25}^{M_{\rm UV,up}} L^1_\nu(z)\frac{L_{\rm
    UV}}{L^1_{\nu_{\rm UV}}(z)}\Phi(M_{\rm UV},z) dM_{\rm UV},
\label{corr}
\end{equation}
where $L^1_\nu(z)$ is the luminosity of the brightest galaxy in the simulation (we assume 
all rare bright galaxies have the same Spectral Energy Distribution (SED) of this one),
$L_{\rm UV}$ is the luminosity corresponding to the UV absolute
magnitude $M_{\rm UV}$.
The wavelength used to calculate the absolute UV magnitude in this paper 
is $1700~\rm \AA$,
$\nu_{\rm UV}$ is the frequency corresponding to this wavelength.
In observations, the selected wavelength corresponding to the UV absolute magnitude 
may be somewhat different in different measurements and 
at different redshifts
\citep{2007ApJ...670..928B,2010ApJ...709L..16O,2010ApJ...709L.133B}, 
however, our results are not sensitive to such differences.  
For the LF $\Phi(M_{\rm UV},z)$ in the redshift range $5 < z < 10$,
we use the Schechter formula \citep{1976ApJ...203..297S} with 
the redshift-dependent parameters given by 
\cite{2011ApJ...737...90B} (see their Sec. 7.5), who fitted the
observed LFs in $z\sim4-8$ and extrapolated them to higher redshifts.
For redshifts above 10,
we simply add an exponential tail normalized to  the simulated
LF amplitude at $M_{\rm UV,up}$. We find that 
this results only a small correction.
As discussed below (see also bottom panel
of Figure \ref{Inu0}), $\sim 90\%$ of the high-$z$ galaxy contribution
to the NIRB flux comes from sources at $5<z<8$  where the correction is at most $12\%$.
This correction is also applied to the calculation 
of ionizing photons below.

For each galaxy, the radiation comes from two different mechanisms: the stellar emission 
and the nebular emission. The former comes directly from the surface of stars, while the latter 
is generated by the ionized nebula around stars and depends on the fraction of ionizing 
photons that cannot escape into the intergalactic medium (IGM), i.e., $1-f_{\rm esc}$,
where $f_{\rm esc}$ is the escape fraction.
Ionizing photons escaping from galaxies would ionize the IGM;
such ionized gas could also produce the nebular emission. However, 
due to the very low recombination rate, 
as shown 
in, e.g., \cite{2001MNRAS.321..593N} and 
\cite{2012ApJ...756...92C}, its emissivity is much weaker than the radiation from galaxies, so 
we ignore this contribution in this paper.
The IGM contribution to the NIRB fluctuations is also negligible
\citep{2010ApJ...710.1089F}.

To determine the escape fraction
averaged over the galaxy populations present at a given redshift, we proceed as follows. First we
  compute the number of ionizing photons
emitted per baryon in collapsed objects as:
\begin{equation}
f_\star N_\gamma=
\frac{\sum_{i=1}^N \left[q^{\rm II}_{\rm H}(\tau^{{\rm II},i},Z^i)\dot{M}^{{\rm II},i}_\star\tau^{{\rm II}
,i}+q^{\rm III}_{\rm H}M^{{\rm III},i}_\star\tau^{\rm III}\right]}{\sum_{i=1}^NM^i_{\rm gas}},
\end{equation}
where  $q^{\rm II}_{\rm H}$ is the emission rate of ionizing photos
from Pop II stars  (this quantity depends on both the age and the
metallicity of the stellar population)  corresponding to a continuous
star formation rate $1~M_\odot\rm yr^{-1}$. We derive this quantity
from the {\tt Starburst99}
templates\footnote{http://www.stsci.edu/science/starburst99/docs/default.htm} \citep{1999ApJS..123....3L,2005ApJ...621..695V,2010ApJS..189..309L}
adopting the mean age, $\tau^{{\rm II},i}$, and metallicity, $Z^i$, of
each simulated galaxy.  $q^{\rm III}_{\rm H}$ is the emission rate of
ionizing photons for Pop III stars
according to \cite{2002A&A...382...28S}
\footnote{Note the different dimensions of $q^{\rm II}_{\rm H}$ and 
$q^{\rm III}_{\rm H}$: the former corresponds to per unit star formation rate while 
the latter corresponds to per unit stellar mass.}.
$M^i_{\rm gas}$ is the gas content,
$\dot{M}^{{\rm II},i}_\star$ is the mean star formation rate of Pop II stars in 
this galaxy, while $M^{{\rm III},i}_\star$ is the cumulative mass of Pop III stars.
We use a mean lifetime $\tau^{\rm III}=2.5\times10^6$~yr for massive Pop III stars
\citep{2002A&A...382...28S,2011MNRAS.414..847S}.

\begin{figure}
\begin{center}
\includegraphics[scale=0.4]{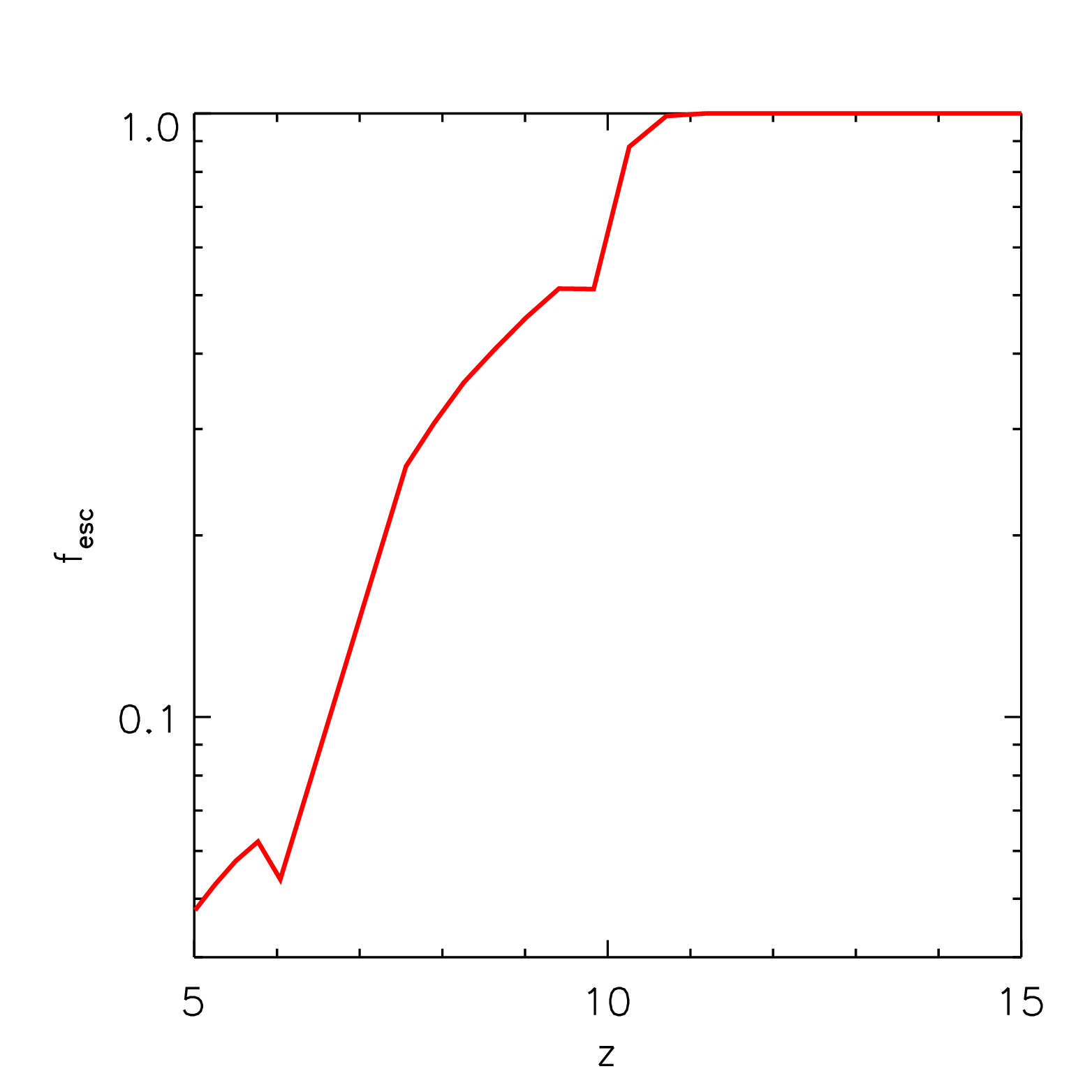}
\caption{Escape fraction evolution from joint LF-reionization constraints.}
\label{fesc}
\end{center}
\end{figure}

We then compare the above quantity with the number of ionizing photons per baryon 
in collapsed objects, $N_{\rm ion}$, required by interpreting the observations as in 
\cite{2012MNRAS.419.1480M} (the ``mean" value) to get the escape fraction,
i.e., $f_{\rm esc}={\rm min}(\frac{CN_{\rm ion}}{f_\star N_\gamma},1.0)$,
where $C$ is the clumping factor. Throughout this paper we assume $C = 1$
to get the minimum $f_{\rm esc}$ therefore the maximum contribution of the 
nebular emission to the NIRB. We note that the clumping factor could 
be higher than 1 even at high redshifts
\citep{2009MNRAS.394.1812P,2012ApJ...747..100S}. For example, \cite{2012ApJ...747..100S}
gives $C\approx 3~(1.7)$ at $z = 5~(9)$ by numerical simulations.
Our nebular emission is therefore reduced by about 10\% to 60\% 
from $z = 5$ to $z = 9$
if this clumping factor is adopted. However, 
for Pop II stars which are the dominant contributors to the NIRB,
the nebular emission is 
much smaller than the stellar emission.
So the final reduction in the NIRB would be much smaller.
Furthermore, we will show later that the flux from high redshift galaxies and Pop III 
stars is unable to explain the observed fluctuations level, so
that a
reduction in the NIRB flux
would in any case strengthen this conclusion.
We plot the derived $f_{\rm esc}$ as a function of redshift in Figure \ref{fesc}.
There is a clear trend of an increasing escape fraction towards higher redshifts;
it reaches 1 at $z \approx 11$. At $z = 5$, the final redshift of the simulation,
$f_{\rm esc}\approx0.05$.
Although required by reionization data, an increasing trend of $f_{\rm esc}(z)$
has not yet been fully understood theoretically in spite of the several, often
conflicting, studies on this problem.

Based on observations,
\cite{2006MNRAS.371L...1I} concluded that $f_{\rm esc} > 0.1$ when $z > 4$;
by combining the observations of 
Lyman $\alpha$ absorption and UV LF, and also using 
N-body simulations and semi-analytical prescriptions to model the ionizing
background, \cite{2010PASA...27..110S} found that for galaxies 
at $z\sim5.5-6$, if the minimum mass of star forming galaxies corresponds to the hydrogen cooling 
threshold, $f_{\rm esc}\sim0.05-0.1$; \cite{2010MNRAS.401.2561W} used the 
star formation rate derived from gamma-ray burst observations
to conclude that in the redshift range $4-8.5$, $f_{\rm esc}\sim0.05$;
\cite{2009ApJ...693..984W}, by radiation hydrodynamical simulations,
found that at redshift 8 for galaxies with $M_{\rm vir} < 10^{7.5}~M_\odot$,
$f_{\rm esc}\sim0.05-0.1$, while for more massive galaxies
$f_{\rm esc}\sim0.4$, if a normal IMF is adopted; also via simulations,
\cite{2010ApJ...710.1239R} found $f_{\rm esc}\sim0.8$ when $z=10$.
The escape fraction derived by us is broadly consistent with these
values. The important difference however is that our derivation of $f_{\rm esc}$
matches both the LF and the reionization history simultaneously, i.e., a more
phenomenological derivation, so that we can get around the detailed physical 
mechanisms of the escape fraction.
Different from our approach,
  \cite{2013MNRAS.428L...1M} computed the   
LFs of high redshift galaxies by means of semi-analytical models and
derived the star formation efficiency $f_\star$ required to match the
observed ones. They found $f_{\rm esc} \approx 0.07$
at $z = 6$ and $f_{\rm esc} \approx 0.16$ at $z = 7$, which are consistent with
our $f_{\rm esc} \approx 0.06~(0.18)$ at those two redshifts. At higher redshifts,
however, their escape fraction is somewhat lower than ours.

As a final remark, we underline that when computing the escape fraction, we do not make a distinction between Pop III and Pop II stars. 
In the calculation of $f_\star N_{\gamma}$ ionizing photons from both populations
are accounted for, $f_{\rm esc}$ can be regarded  
as a kind of ``effective" escape fraction averaged over the galaxy population. In principle, $f_{\rm esc}$  for Pop III stars should be 
higher due to their harder spectrum. However, as we will see in Section \ref{results}, Pop III stars only contribute a negligible 
flux to the present-day NIRB, a more detailed modeling is then not necessary.

The luminosity of the $i$-th galaxy, $L^i_\nu$, is the sum of the
contribution of Pop II and Pop III stars. For Pop II stars we use the age
and metallicity dependent spectrum templates provided by the
{\tt Starburst99} code. The nebular emission contribution has been
renormalized by adopting the escape fraction computed above.
In addition to the free-free, 
free-bound and two-photon emissions which have already been  
included in {\tt Starburst99}, we add the Lyman $\alpha$ emission 
to the template by using \citep{2006ApJ...646..703F}
\begin{equation}
l_{\alpha}(\nu,\tau^{{\rm II},i},Z^i,z)=f_{\alpha}h_{\rm p}\nu_{\alpha}\phi(\nu-\nu_\alpha)q^{{\rm II}}_{\rm H}( \tau^{{\rm II},i},Z^i)[1-f_{\rm esc}(z)],
\label{lya}
\end{equation}
in which $f_\alpha=0.64$ \citep{2006ApJ...646..703F},
$h_{\rm p}$ is the Plank constant, $\nu_\alpha=2.47\times10^{15}$~Hz
is the frequency of Lyman 
$\alpha$ photons.
We use the line profile $\phi(\nu-\nu_\alpha)$ 
provided in \cite{2002MNRAS.336.1082S}:
\begin {equation}
\phi(\nu-\nu_\alpha)= \left\{
\begin{tabular}{ll}
$\nu_\star(z)(\nu-\nu_\alpha)^2{\rm exp}[-\nu_\star(z)/|\nu-\nu_\alpha|]$~if $\nu \le \nu_\alpha$ \\
0~~~~~~~~~~~~~~~~~~~~~~~~~~~~~~~~~~~~~~~~~~~~~~if $\nu > \nu_\alpha$,
\end{tabular}
\right.
\end{equation}
where
\begin{equation}
\nu_\star(z)=1.5\times10^{11}\left(\frac{\Omega_bh^2}{0.019}\right)\left(\frac{h}{0.7}\right)^{-1}\frac{(1+z)^3}{\sqrt{\Omega_m(1+z)^3+\Omega_\Lambda}}~\rm Hz
\end{equation}
is the fitted form of results given in \cite{1999ApJ...524..527L}.

For the template of Pop III stars, $l_\nu^{\rm III}$, we still use the spectrum in 
\cite{2002A&A...382...28S}, but renormalize the nebular emission part by the 
factor $1-f_{\rm esc}$. The luminosity of the $i$-th galaxy is then given by 
\citep{2011MNRAS.414..847S}
\begin{equation}
L^i_\nu(z)=l_\nu^{{\rm II}}(\tau^{{\rm II},i},Z^i,z)\dot{M}^{{\rm II},i}_\star+
l_\nu^{{\rm III}}(z)\dot{M}_\star^{{\rm III},i}\tau^{{\rm III}},
\end{equation}
here the Lyman $\alpha$ emission in Eq. (\ref{lya}) has already been included in 
$l_\nu^{{\rm II}}$.

\begin{figure}
\begin{center}
\includegraphics[scale=0.4]{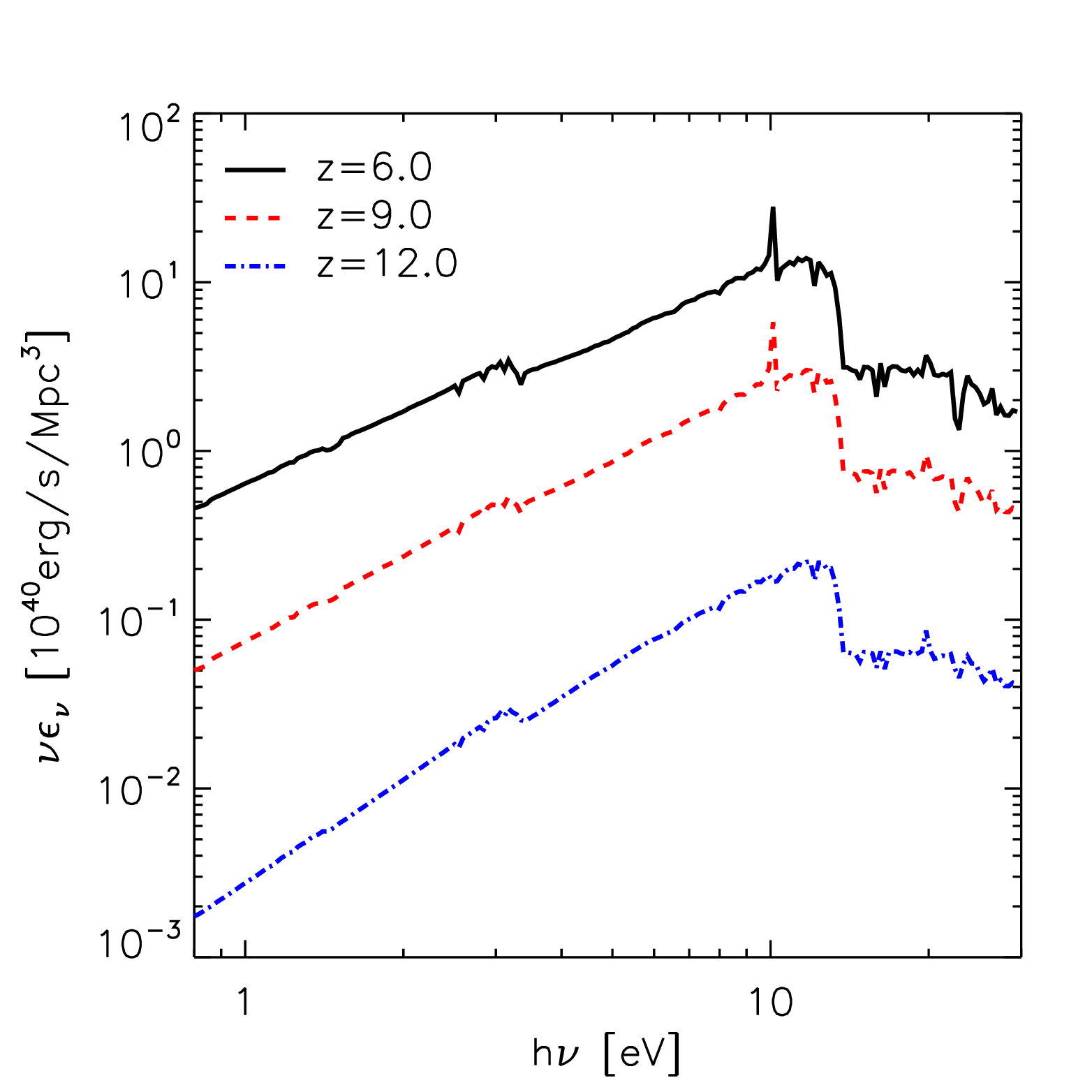}
\caption{The emissivity of simulated galaxies 
at redshift 12.0 (dash-dotted), 9.0 (dashed) and 6.0
(solid) respectively.}
\label{emiss}
\end{center}
\end{figure}

With the luminosity for each galaxy given as above,
we can then obtain the emissivity according to
Eq. (\ref{e}).
As an example, we plot the $\nu\epsilon(\nu,z)$ at 
redshifts 12.0, 9.0 and 6.0 respectively in Figure 
\ref{emiss}. 
At high redshifts, the escape fraction $\approx1.0$, yielding a very weak 
Ly$\alpha$ line, since such emission is produced by 
recombinations of the ionized nebula around stars.
At lower redshifts, the escape fraction drops, while more ionizing photons
are absorbed by the material around the stars, and producing more
Ly$\alpha$ emission which  
is more clearly seen in the spectrum.

The part of spectrum with energy below 10.2 eV is of the most interest to us,
here the spectrum becomes increasingly flatter at later time. For example, at $z = 12$, the slope of 
$\nu\epsilon(\nu,z) \propto \nu^\beta$ with $\beta \approx 2$, while at $z = 5$
$\beta \approx 1.2$. This is clearly the result of an aging effect enhancing the rest frame
optical/IR bands flux with respect to the UV ones. Since the NIRB
from $z > 5$ galaxies 
is dominated by the 
lower redshift galaxies ($5 < z < 8$), we do not expect to have a very steep NIRB spectrum, as
we will see in the results presented in Sec. \ref{results}.

\subsection{NIRB Fluctuations}
Using the Limber approximation, 
the angular power spectrum of the fluctuations of the flux field 
is \citep{2012ApJ...756...92C}
\begin{equation}
C_l=\int_{z_{\rm min}}^{z_{\rm max}}\frac{dz}{r^2(z)(1+z)^4}\frac{dr}{dz}[\nu\epsilon(\nu,z){\rm e}^{-\tau_{\rm eff}(\nu_0,z)}]^2P_{\rm gg}(k,z),
\label{angpow}
\end{equation}
where $r(z)$ is the comoving distance and 
$P_{\rm gg}(k,z)$ is the galaxy-galaxy power spectrum, $k=l/r(z)$. 
In Eq. (\ref{angpow}) we assume that the luminous properties of galaxies are independent of 
their locations, so that the only factor which determines their contribution 
to the NIRB fluctuations
is their spatial fluctuations (see \citealt{2012MNRAS.421.2832S} for an improved model). 

The $10~h^{-1}$~Mpc box size of \cite{2011MNRAS.414..847S} simulations 
is too small to provide us with the large scale correlation function of galaxies
(for sources at $z = 6$, the comoving transverse separation corresponding to $1^{\ensuremath{\circ}}$
angular size 
is about $100~h^{-1}$Mpc), so we 
use the halo model \citep{2002PhR...372....1C,2004MNRAS.348..250C}
to calculate the galaxy-galaxy power spectrum.
This power spectrum is composed of two parts, the one-halo term from the 
correlation of galaxies in the same halo
(including the central galaxies and satellite galaxies),
and the two-halo term from galaxies in different halos:
\begin{equation}
P_{\rm gg}(k,z)=P^{1h}_{\rm gg}(k,z)+P^{2h}_{\rm gg}(k,z).
\end{equation}
Assuming that the distribution of galaxies in a halo traces the 
profile of dark matter, and the mean number of central galaxies
and satellite galaxies in a halo with mass $M$ are 
$\langle N_{\rm sat}\rangle$ and $\langle N_{\rm cen}\rangle$, respectively,
we have
\begin{equation}
P^{1h}_{\rm gg}(k,z)=
\int_{M_{\rm min}(z)}^{M_{\rm max}(z)}dM\frac{dn}{dM}
\frac{2\langle N_{\rm sat}\rangle \langle N_{\rm cen}\rangle u(M,k)+
\langle N_{\rm cen}\rangle^2 u^2(M,k)}{\bar{n}^2_{\rm gal}},
\end{equation}
and   
\begin{equation}
P^{2h}_{\rm gg}(k,z)=P_{\rm lin}(k,z)
\times\left[\int_{M_{\rm min}(z)}^{M_{\rm max}(z)} dM\frac{dn}{dM}b(M,z)
\frac{\langle N_{\rm sat}\rangle+\langle N_{\rm cen} \rangle}{\bar{n}_{\rm gal}}u(M,k)\right]^2.
\end{equation}
In the above expressions, $M_{\rm min}(z)$ is the minimum mass of halos that could host 
galaxies, and we set it to be the minimum mass of halos that contain stars
in our simulations, which is $\sim(2-8)\times10^7~M_\odot$, depending on the redshift.
$M_{\rm max}(z)$ is the maximum mass contributing to the emissivity and the clustering,
we will describe how to determine it later,
${dn}/{dM}$ is the mass function \citep{1999MNRAS.308..119S,2001MNRAS.323....1S},
while $u(M,k)$ is the normalized Fourier transform of the halo profile. For 
a NFW profile \citep{1997ApJ...490..493N}, the analytical expression is given by 
\cite{2002PhR...372....1C}, and we use the concentration parameter, $c_{\rm M}$,
from \cite{2012MNRAS.423.3018P} which fits simulations well. 
However, we find that our results are insensitive to the use of different concentrations, as
e.g., from \cite{2011ApJ...736...59Z}. Even if a
very different concentration parameter
is adopted, its impacts are non-negligible only in the one-halo term which dominates 
the signal at small scales, where galaxy clustering is well below the shot noise. 
As we are interested primarily in the large-scale ($> 1'$) clustering,
our conclusions are unaffected by the adopted value of $c_{\rm M}$. Finally,
for the halo bias $b(M,z)$
we use the formula and fitted parameters given by \cite{2010ApJ...724..878T},
which is higher than \cite{2001MNRAS.323....1S}
for massive halos, but is better fit to simulations. 
The linear matter power spectrum $P_{\rm lin}(k,z)$ is taken from 
\cite{1998ApJ...496..605E}. 

The mean number of central galaxies and satellites in a halo with mass 
$M$ is modeled by the halo occupation distribution (HOD) model
\citep{2005ApJ...633..791Z},
\begin{equation}
\langle N_{\rm cen} \rangle =\frac{1}{2}\left[1+{\rm erf}
\left(\frac{{\rm log_{10}}M-{\rm log_{10}}M_{\rm min} }{\sigma_{{\rm log_{10}}M}}\right)\right],
\end{equation}
and
\begin{equation}
\langle N_{\rm sat} \rangle=\frac{1}{2}\left[1+{\rm erf}
\left(\frac{{\rm log_{10}}M-{\rm log_{10}}2M_{\rm min} }{\sigma_{{\rm log_{10}}M}}\right)\right]
\left(\frac{M}{M_{\rm sat}}\right)^{\alpha_s}. 
\end{equation}
We adopt the parameters $M_{\rm sat}=15M_{\rm min}$, $\sigma_{{\rm log_{10}}M}=0.2$ and
$\alpha_s=1.0$, which are from both simulations and semi-analytical models
\citep{2005ApJ...633..791Z},
and observations \citep{2011ApJ...736...59Z}.
With the mean number of central and satellite galaxies in each 
halo, the galaxy number density is simply 
\begin{equation}
\bar{n}_{\rm gal}=\int_{M_{\rm min}(z)}^{M_{\rm max}(z)} dM\frac{dn}{dM}\left( \langle N_{\rm cen}\rangle+\langle
N_{\rm sat} \rangle \right).
\end{equation}

In addition to the above galaxy clustering, 
Poisson fluctuations in the number 
of galaxies would generate shot noise in observations, whose power spectrum 
dominates at small scales.
If the redshift derivative of the number of sources with flux between 
$S$ and $S+dS$ is $\frac{d^2N}{dSdz}$, 
the angular power spectrum of such shot noise is 
\begin{equation}
C^{\rm SN}_l=\frac{1}{\Delta \Omega}\int dz \int dSS^2\frac{d^2N}{dSdz}=
\frac{1}{\Delta \Omega}\int dz \int dMS^2\frac{d^2N}{dMdz},
\end{equation}
where $\Delta \Omega$ is the beam angle.
Considering that 
\begin{equation}
\frac{d^2N}{dMdz}=\frac{dn}{dM}\Delta \Omega r^2\frac{dr}{dz},
\end{equation}
and 
\begin{equation}
S=\frac{L_{\nu}(M){\rm e}^{-\tau_{\rm eff}(\nu_0,z)}}{4\pi r^2(1+z)},
\end{equation}
where $L_{\nu}(M)$ is the luminosity of halos with mass $M$,
the shot noise power spectrum is 
\begin{equation}
C^{\rm SN}_l=\int dz\frac{{\rm e}^{-\tau_{\rm eff}(\nu_0,z)}}{r^2(1+z)^2}\frac{dr}{dz} \int \left[\frac{L_{\nu}(M)}{4\pi M}\right]^2 M^2\frac{dn}{dM}dM. 
\end{equation}
As we assume the luminous properties of galaxies are independent of their location, 
in the square bracket we can simply use an average light-to-mass ratio that is independent of the 
halo mass,
\begin{equation}
\frac{1}{4\pi}\left.\int L_{\nu}(M)\frac{dn}{dM}dM\middle/
\int M\frac{dn}{dM}dM\right.=\frac{\epsilon(\nu,z)}{\rho_h}.
\end{equation}
We finally obtain the shot noise angular power spectrum
\begin{equation}
C^{SN}_l=\int_{z_{\rm min}}^{z_{\rm max}}\frac{cdz}{H(z)r^2(z)(1+z)^4}P^{\rm SN}(z),
\end{equation}
where
\begin{equation}
P^{SN}(z)=\left[\frac{\nu\epsilon(\nu,z){\rm e}^{-\tau_{eff}}}
{\rho_{h}}\right]^2\int_{M_{\rm min}(z)}^{M_{\rm max}(z)} dMM^2\frac{dn}{dM},
\end{equation}
and the halo mass density is $$ \rho_h=\int_{M_{\rm min}(z)}^{M_{\rm max}(z)} dM M\frac{dn}{dM}$$.

In observations, the detected sources are generally removed down to a certain
limiting magnitude, $m_{\rm lim}$, 
the residue is the source-subtracted NIRB fluctuations.
To simulate this, we also remove bright galaxies in the
simulation box and in the bright-end. In theoretical calculations,
this limiting magnitude is determined by letting the predicted
shot noise level match the values found in the measurements.

The apparent limiting magnitude (at wavelength $\lambda_0$) 
is converted into the rest frame absolute magnitude (at wavelength $\lambda_0/(1+z)$)  
$M_{\lambda_0/(1+z)}$ by
\begin{equation}
M_{\lambda_0/(1+z)}=m_{\rm lim}-DM(z)+2.5 {\rm log_{10}}(1+z),
\end{equation}
where $DM(z)$ is the distance modulus \citep{2012ApJ...752..113H}. 
By using a light-to-mass ratio constructed from the simulation, we determine 
the maximum halo mass $M_{\rm max}(z)$.
Things are slightly more complicated when calculating the absolute flux and the spectrum of 
fluctuations, since they both depend on wavelength, 
while the limiting magnitude in observations at different wavelength is different.
In this case we simply give the theoretical prediction without 
removing any sources in the simulation box and the bright-end, as shown in 
Figure \ref{Inu0} and Figure \ref{Cl_spectrum}, corresponding to $M_{\rm max} = \infty$.
Then we 
discuss the effects of galaxy removal, i.e., in Figure \ref{subtract}.
Throughout this paper we adopt $z_{\rm min} = 5$
and $z_{\rm max} = 19$ unless otherwise specified.

\section{RESULTS}\label{results}

\begin{figure}
\begin{center}
\subfigure{\includegraphics[scale=0.4]{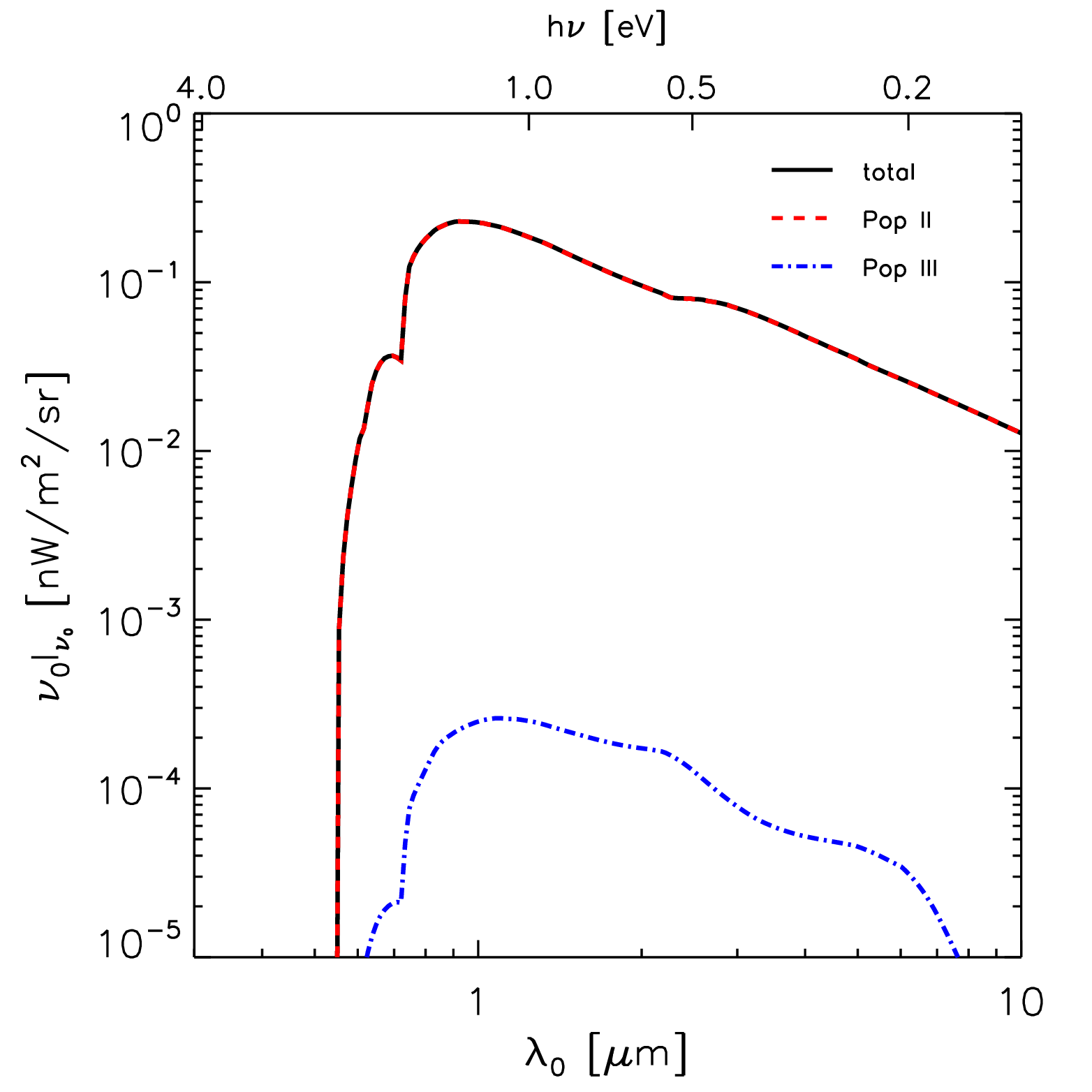}}
\subfigure{\includegraphics[scale=0.4]{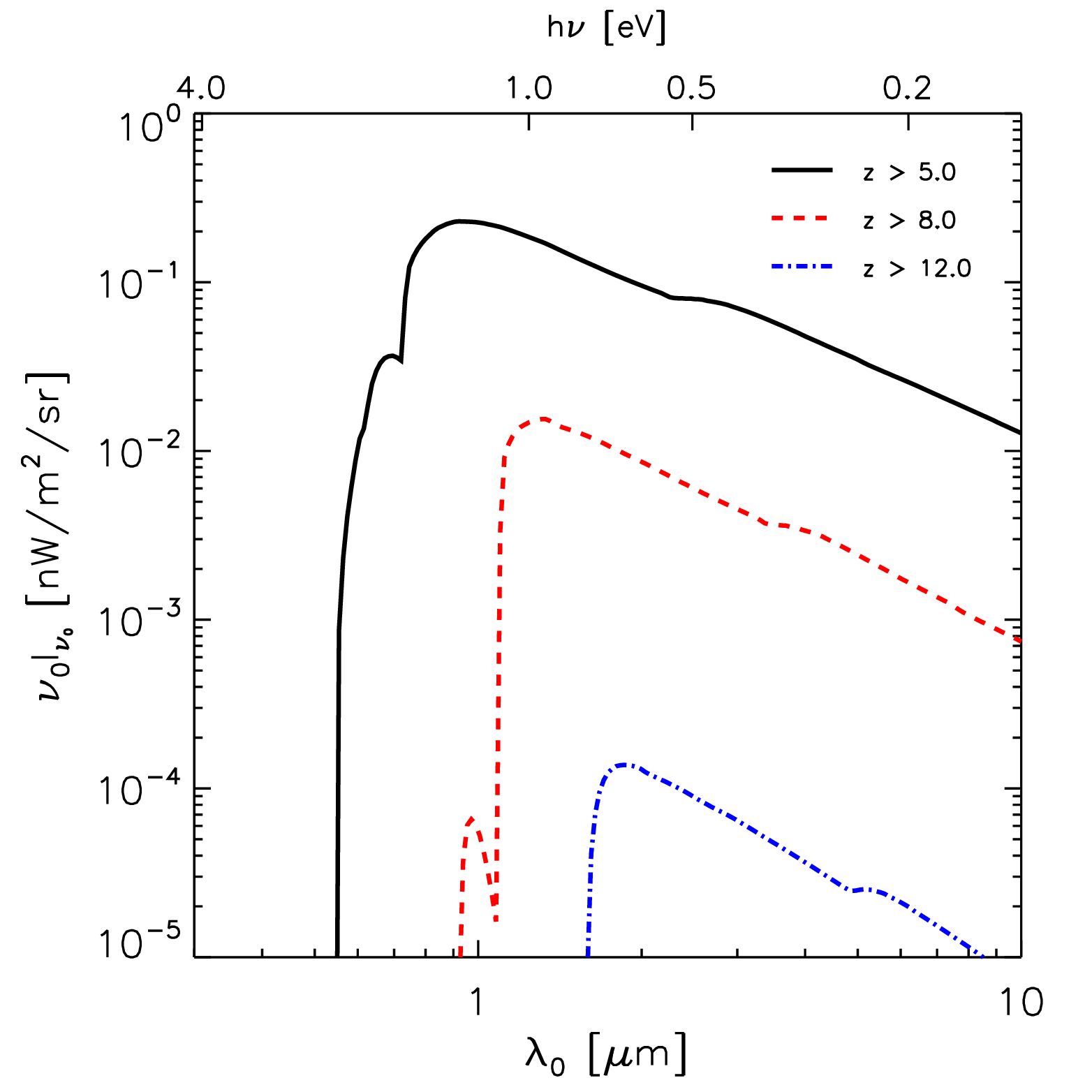}} 
\caption{The NIRB flux from high-$z$ galaxies,
$\nu_0I_{\nu_0}$, as a function of wavelength in the observer frame. 
{\it Left panel:}  Contributions from Pop II (dashed line), Pop III (dash-dotted) stars,  and their sum.
Since the contribution from Pop III stars is very small, the solid line and the dashed line are almost identical.
{\it Right panel:} Contributions from sources in different redshift ranges: $z > 5$ (solid line), $z > 8$ (dashed) and 
$z > 12$ (dash-dotted).
}
\label{Inu0}
\end{center}
\end{figure}

\begin{figure}
\begin{center}
\includegraphics[scale=0.4]{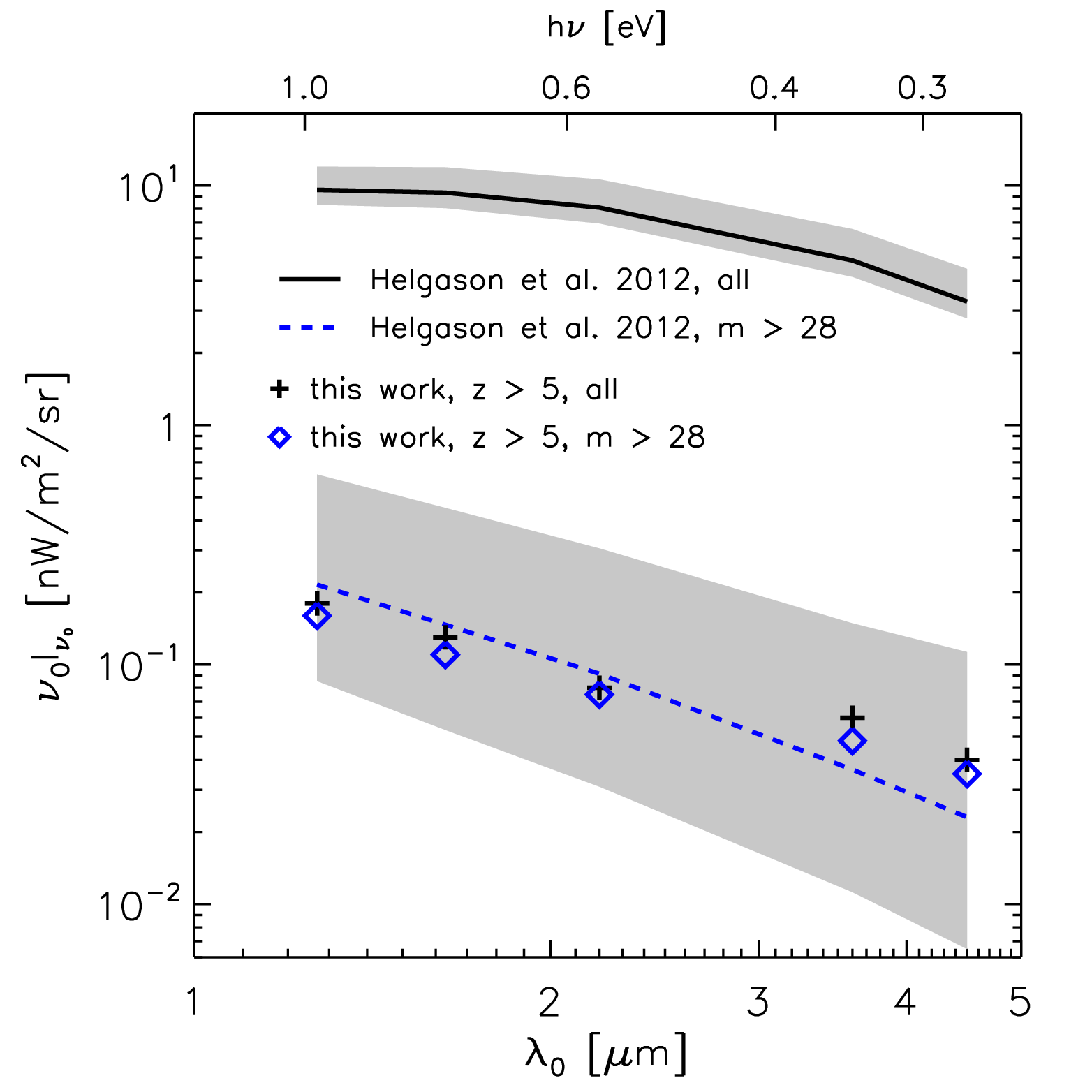}
\caption{NIRB flux in the 1.25~$\rm \mu m$ to 4.5~$\rm \mu m$ wavelength range. The solid line is the 
flux from all galaxies from the ``default" model in \citet{2012ApJ...752..113H}, while the dashed line is the 
remaining flux after removal of all sources brighter than $m_{\rm lim} = 28$. The grey regions
refer to the flux range between the ``HFE" and ``LFE" models in \citet{2012ApJ...752..113H}. 
As a comparison, we plot the flux from all galaxies with $z > 5$ in our work (crosses), and the flux after removal of sources down to
$m_{\rm lim} = 28$ (diamonds). Before any galaxy removal, the flux from galaxies 
with $z > 5$ is only a few percent of the overall flux in \citet{2012ApJ...752..113H},
i.e. the low-$z$ galaxies dominate. However, after removal of galaxies with $m_{\rm lim} < 28$,
the flux from the remaining galaxies at all redshifts in \citet{2012ApJ...752..113H} is
comparable with that from the remaining galaxies with $z > 5$ in our work.
}
\label{subtract}
\end{center}
\end{figure}

We start by presenting the 
contribution of high-$z$ galaxies to the absolute flux of the NIRB 
observed at $z = 0$ in the (observer frame) wavelength range
$0.3-10$~$\rm \mu m$.
Figure \ref{Inu0} (top panel) shows the predicted cumulative flux  
when all sources with $z > 5$ are included; also shown separately are the contributions from Pop II  
and Pop III stars. The flux peak value is $\rm 0.2~nW/m^2/sr$ at $\lambda_0=0.9~\rm \mu m$, and decreases to
$\rm 0.04~nW/m^2/sr$ at $\lambda_0=4.5~\rm \mu m$. The small bump on the left side of the peak is due to 
intergalactic Ly$\alpha$ absorption by intervening neutral hydrogen. 

We find that in our case, the Pop III contribution is almost negligible (it never exceeds 1\%). 
This is not surprising, for in the simulation the Pop III star formation rate is about three 
orders of magnitude lower than that of Pop II stars at $z=10$; the ratio is even smaller below this redshift 
\citep{2007MNRAS.382..945T}. Stated differently, halos with the highest Pop III stellar fraction are usually 
smaller and less luminous, and their contribution to the total luminosity is very low \citep{2011MNRAS.414..847S}. 
This means that it is very  
difficult to find Pop III signatures by means of NIRB observations. 

In the bottom panel of Figure \ref{Inu0}, we plot the contributions from 
the sources above redshift 5.0, 8.0 and 12.0, respectively. 
From the figure, it is clear that the contributions from the sources at $5 < z < 8$ dominate,
providing about 90\% of the flux from all sources with
$z > 5$. Most of these sources are the low-luminosity galaxies which cannot be detected individually in current surveys, and they are
believed to be the major contributors to reionization. In principle, then, the NIRB could be a perfect tool to study the reionization 
sources without detecting them individually.

One way to approach the high-$z$ components is to remove bright sources in the field of view.
We plot the flux before and after removal of bright galaxies in Figure \ref{subtract} at wavelengths from 
1.25~$\rm \mu m$ to 4.5~$\rm \mu m$, corresponding to the $J$ through $M$ bands.
The crosses refer to flux from all galaxies with $z > 5$ in our work, while the solid line corresponds
to the flux from 
all galaxies at all redshifts in the ``default" model of \cite{2012ApJ...752..113H}. After removal of galaxies brighter than $m_{\rm lim} = 28$,
the flux from $z > 5$ galaxies in our work is shown by diamonds, while flux from remaining galaxies in 
\cite{2012ApJ...752..113H} is shown by the dashed line. 
The flux from all galaxies (solid line) is about 1-2 orders of magnitude 
larger than that from galaxies with $z > 5$ (crosses) in our work. Hence, without bright galaxy removal,  
$z<5$ galaxies largely dominate the NIRB flux. However, if we remove the galaxies down to $m_{\rm lim} = 28$, the 
flux from the remaining galaxies at all redshifts in \cite{2012ApJ...752..113H}
(dashed line) is comparable to that from the remaining 
galaxies with $z > 5$ in our work. Even considering the uncertainties on the faint-end of LFs
(the shaded regions), in the source-subtracted flux,
galaxies at $z>5$ still contribute at least $\sim$20\%$-$30\%
of the flux from galaxies at all redshifts and
fainter than $m = 28$. So at least in principle we can access the
signal of reionization sources by subtracting the bright galaxies
from the NIRB.

Before moving to fluctuations, we emphasize that the expected contribution of high-$z$ galaxies (including Pop III stars) to the NIRB
flux is very small compared with the residual flux in the measurement of \cite{2005ApJ...626...31M}. These sources largely fall short 
of accounting for such residual ($\sim60-6~\rm nW/m^2/sr$ in the wavelength range $1.4-4~\rm \mu m$).
It has to be reminded that the flux measured by Matsumoto et al. is likely to be still dominated by incomplete zodiacal light subtraction, as discussed by   \cite{2007ApJ...657..669T}, who concluded that no residual flux is present.
Considering the difficulties in modeling the zodiacal light accurately, currently the residual flux 
measurements are not very useful to constrain models. Therefore, we will base all the conclusions in the present paper
on the analysis of fluctuations only, see below.

\begin{figure*}
\begin{center}
\subfigure{\includegraphics[scale=0.4]{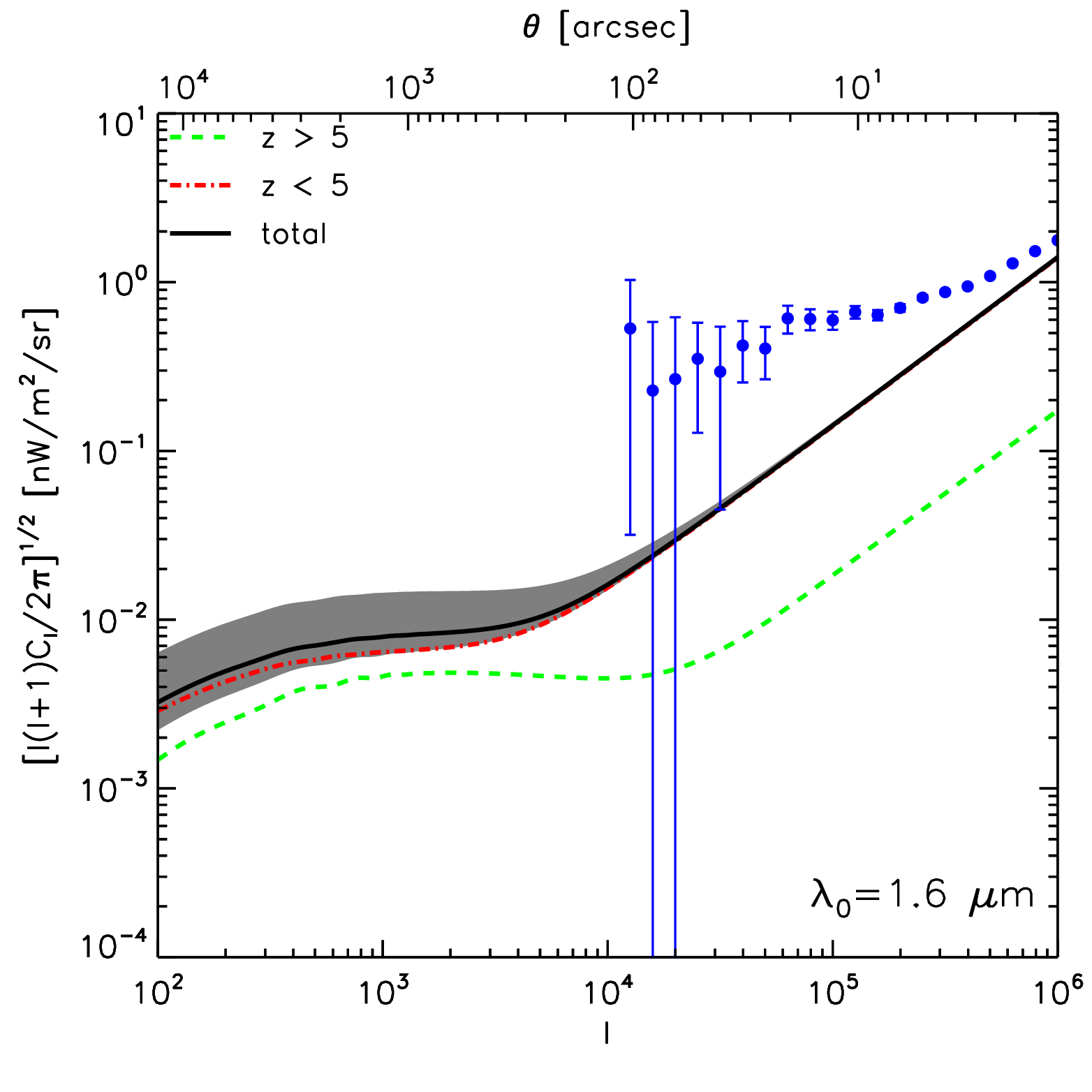}}
\subfigure{\includegraphics[scale=0.4]{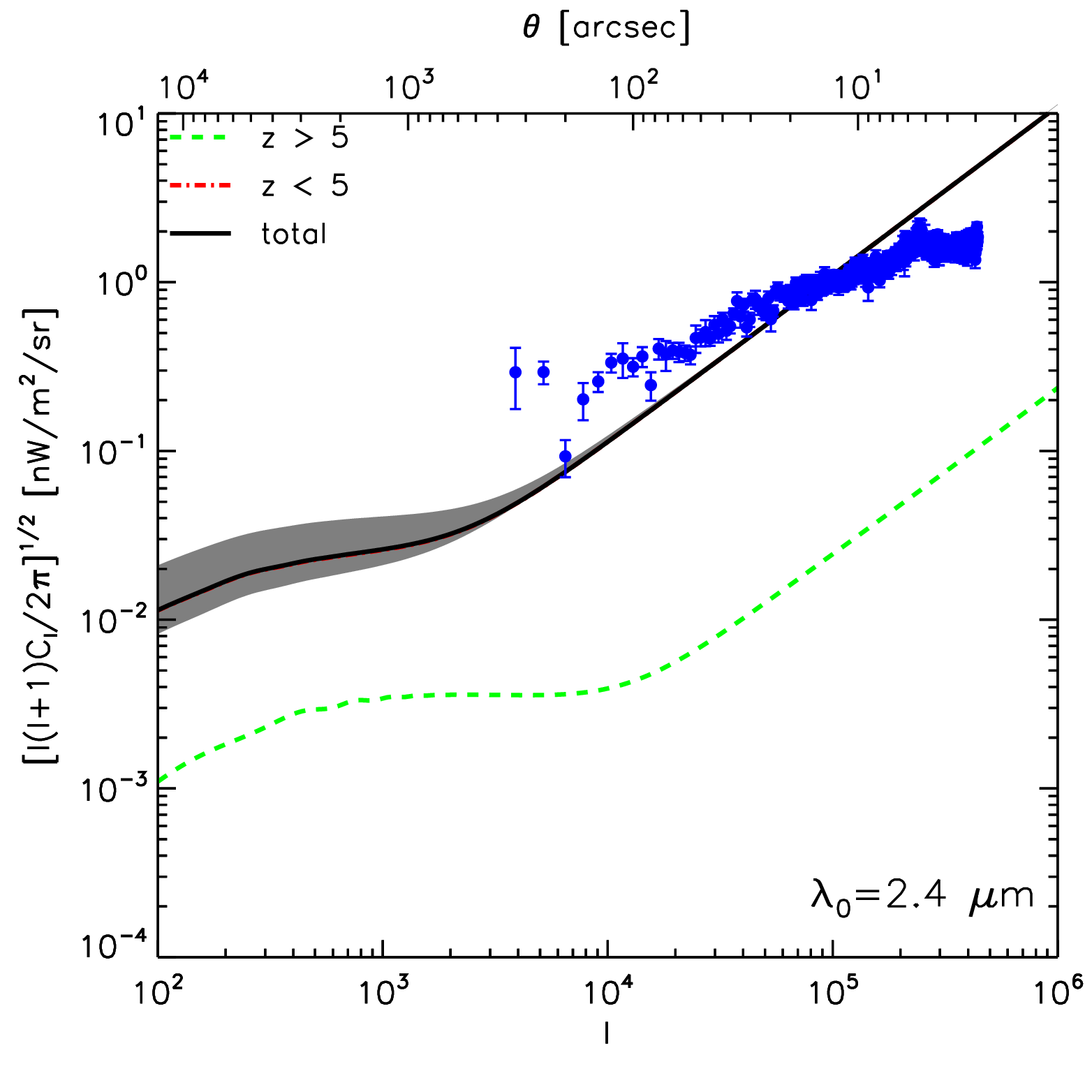}}
\subfigure{\includegraphics[scale=0.4]{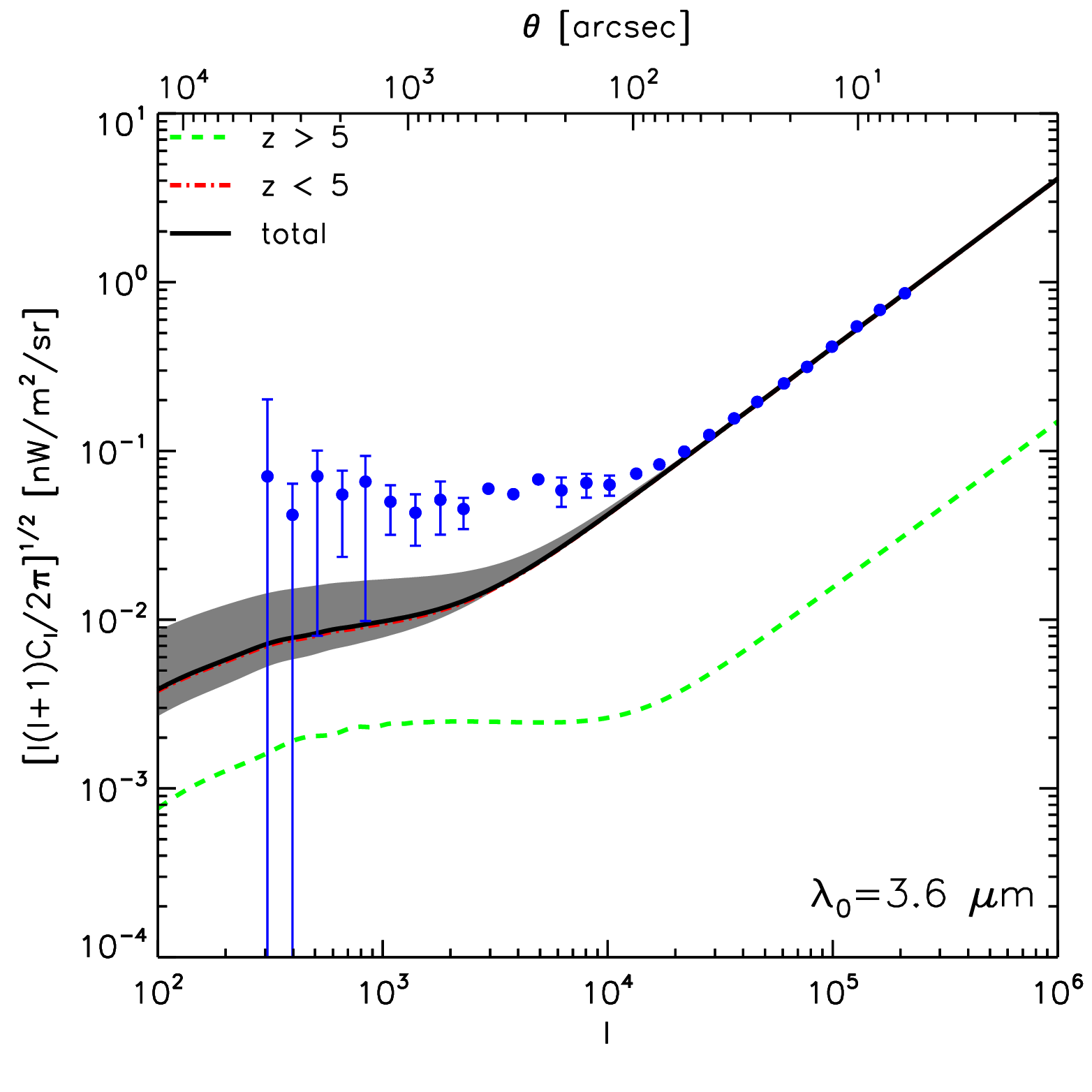}}
\subfigure{\includegraphics[scale=0.4]{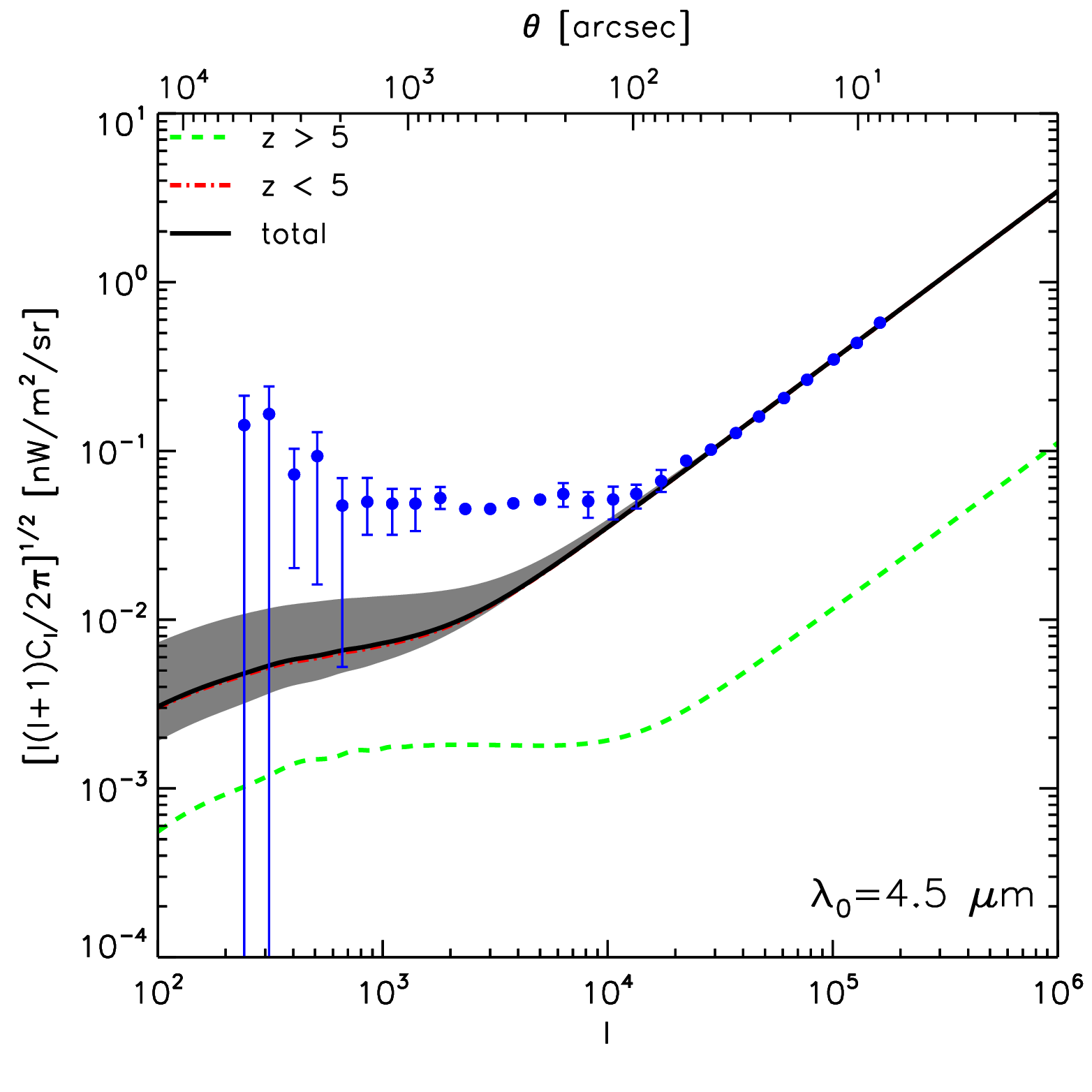}}
\caption{NIRB fluctuations angular power spectrum at different (observer frame) wavelengths, 
as labeled in each panel, both the galaxy clustering and the shot noise are included.
The dashed lines represent the contribution from high-$z$ galaxies studied in this work,
while the dash-dotted lines represent the contribution from low-$z$ galaxies 
reconstructed by \citet{2012ApJ...752..113H} (their ``default"
model). The solid lines are the sum of these.
Note that in all panels except the upper left one where 
the limiting magnitude is very faint, the 
dash-dotted line and the solid line are almost identical.
The shaded regions are the range of the total power spectrum when 
considering the different faint end of the LFs which are likely to 
bracket the real case (the ``HFE" and ``LFE" models in \citealt{2012ApJ...752..113H}).
We also plot the observations
at wavelength $1.6~\rm \mu m$ \citep{2007ApJ...657..669T}, $2.4~\rm \mu m$ \citep{2011ApJ...742..124M},
$3.6~\rm \mu m$ and $4.5~\rm \mu m$ (\citealt{2012Natur.490..514C}, which agrees well with another 
recent measurements, i.e., \citealt{2012ApJ...753...63K}, but extends to larger angular scales)
by filled circles with errorbars. In all our theoretical predictions we have already removed the 
bright sources to reach the shot noise level that matches each measurement, see text. 
}
\label{Cl}
\end{center}
\end{figure*}

The fluctuations of the NIRB after subtracting galaxies down to the detection limits of observations,
$\sqrt{l(l+1)C_l/(2\pi)}$, at $\lambda_0=1.6,~2.4,~3.6,~4.5~\rm \mu m$ are shown by the thick solid 
line in each panel of Figure \ref{Cl}. The contribution from $z > 5$ faint galaxies, which is studied in 
this work, and the contribution from $z < 5$ galaxies, which is calculated by following 
the reconstruction of \cite{2012ApJ...752..113H}, are shown by dashed line and dash-dotted line 
respectively.
Unless the limiting magnitude is very faint so that  the 
relative fraction of the contribution of high-$z$ galaxies is larger
as in the upper left panel, the contribution of 
$z > 5$ galaxies is negligible compared with the 
$z < 5$ galaxies, i.e., the total amplitude (solid line) coincides with
that of $z < 5$ galaxies (dash-dotted line).
To account for the uncertainties 
of the faint-end of LFs, \cite{2012ApJ...752..113H} considered two models of the faint end of LFs 
(adopted for low-$z$ faint galaxies here) which are 
likely to bracket the real case. Considering this we show the range of the total power spectrum 
by shaded regions.
We also plot observations at corresponding wavelength in each panel by 
filled circles with errorbars, which are from 
\cite{2007ApJ...657..669T} ($1.6~\rm \mu m$), \cite{2011ApJ...742..124M} ($2.4~\rm \mu m$),
\cite{2012Natur.490..514C} (3.6 and 4.5 $\rm \mu m$) respectively. 
The measurements of \cite{2012Natur.490..514C} agree well with observations of 
\cite{2012ApJ...753...63K} at the same wavelength, but extend to larger angular scales. 
In the theoretical predictions, we remove the bright sources by selecting a limiting magnitude 
at each wavelength to get a shot noise level of the 
remaining fainter galaxies (including both low-$z$ and high-$z$ ones, but the latter is 
almost negligible) that matches each measurement,
i.e., $m_{\rm lim}=$ 26.7, 23.2, 23.9 and 23.8 
respectively, the first two values are the same as 
\cite{2012ApJ...752..113H}. 

At small scales where the shot noise 
dominates, the model predictions should match the observations, as shown in the 
3.6 and 4.5 $\rm \mu m$ panels. In the 2.4~$\rm \mu m$ panel, there is some 
discrepancy at small scales, this is because the suppression 
of the power by beam effects is not corrected in the observations data. 
For the 1.6~$\rm \mu m$ case, a footnote in \cite{2012ApJ...752..113H}
noted that the images at other wavelength are used to subtract bright sources, 
so there would be spread on the limiting magnitudes.

From the figure, we also see that the contributions of the $z > 5$ galaxies (dashed lines)
exceed the shot noise level on large angular scales ($l < 10^4$), this means that the 
NIRB fluctuations do have the potential to provide us information on the nature of the undetected 
reionization sources.
However, the predicted amplitudes  
are only $\sim (2-4) \times10^{-3}~\rm nW/m^2/sr$,
which is even much smaller 
than the contribution from low-$z$ faint galaxies,
and both the high-$z$ and low-$z$ contributions are much smaller than
the observed values,  which are at the $\sim 0.1~\rm nW/m^2/sr$ level.
Even considering the uncertainties about the faint-end of LFs,
the difference is still quite significant,  
again indicates the existence of one or more unknown component(s) we are missing.
Somewhat surprisingly but interestingly, 
the missing component has a clustering signal very similar to that of
the high redshift galaxies, and extends to degree angular scales. 
Obviously, this component/foreground must be identified before
we can make further progress 
and use the NIRB to study reionization sources.

\begin{figure}
\begin{center}
\includegraphics[scale=0.4]{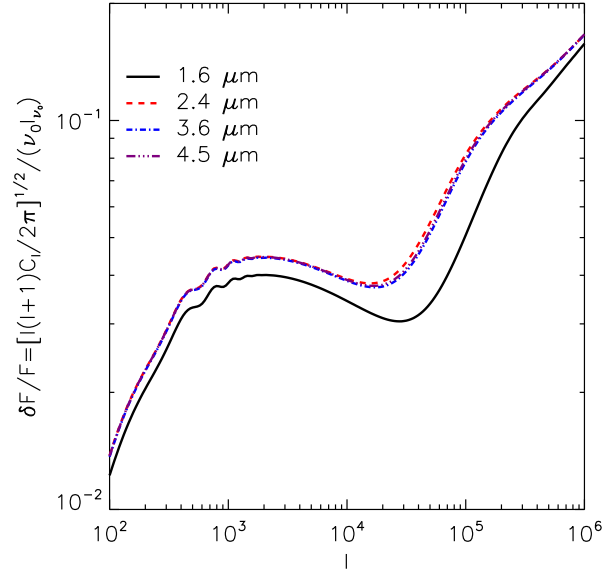}
\caption{The ratio of $\delta F/F$ as a function of $l$ for four different wavelengths.}
\label{deltaF_F}
\end{center}
\end{figure}

Next, we define 
the fluctuation amplitude $\delta F=\sqrt{l(l+1)C_l/(2\pi)}$,
and plot its ratio to $F=\nu_0I_{\nu_0}$ in Figure \ref{deltaF_F}. 
Such relative fluctuation is almost independent of $\lambda_0$ (see also \citealt{2010ApJ...710.1089F}), 
$\delta F/F \sim$ 4\% at $l=2000$, with the only slight deviation of the $1.6~\rm\mu m$ band where it is somewhat lower than in 
redder bands, as a consequence of a deeper ($m_{\rm lim} \approx 27$) galaxy removal. Nicely,  the relative fluctuation agrees with that 
found by \cite{2010ApJ...710.1089F} and \cite{2012ApJ...756...92C}. In addition, $\delta F/F$ increases with $z_{\rm min}$, i.e., 
high redshift sources have higher relative fluctuations. For example, $\delta F/F= 7$\% for $z_{\rm min}=8.0$, while it reaches 12\% if 
$z_{\rm min}=12.0$ is adopted. It reflects the more biased spatial distributions 
of higher redshift sources.
The relative fluctuation $\delta F/F$ is only weakly dependent on the intrinsic properties 
of galaxies \citep{2010ApJ...710.1089F}, but more so on the spatial clustering features.
Thus, $\delta F/F$ is a key indicator to identify NIRB sources; yet, in practice, it is hard to get an accurate absolute flux.

\begin{figure}
\begin{center}
\includegraphics[scale=0.4]{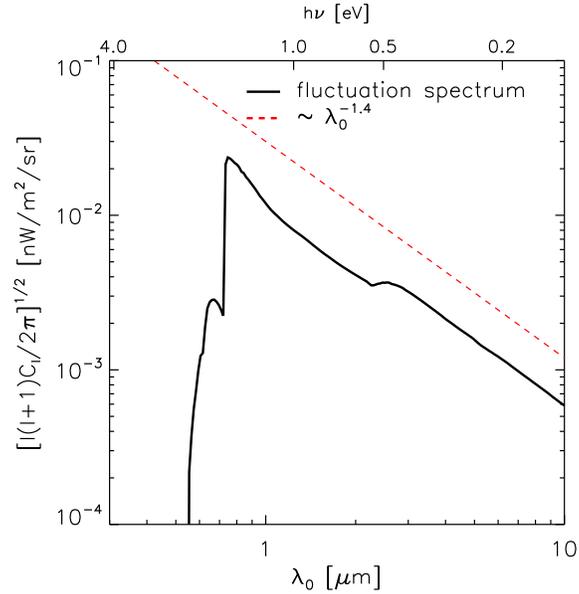}
\caption{The spectrum of the NIRB (contributed by all
$z > 5$ galaxies) fluctuations (solid line) at $l=2000$, the dashed line shows a 
$\lambda_0^{-1.4}$ law.}
\label{Cl_spectrum}
\end{center}
\end{figure}

The spectrum of the fluctuations, $\delta F(\lambda_0)$ from all galaxies with $z > 5$ at $l=2000$, shown in 
Figure \ref{Cl_spectrum}, has a slope $\lambda_0^p$, with $p = -1.4$ above 1~$\mu$m. Such slope is essentially the same
as that of the flux, reflecting the above mentioned wavelength independency of $\delta F/F$.

\section{CONCLUSIONS}\label{conclusions}

By combining high resolution cosmological N-body/hydrodynamical simulations and an 
analytical model, we predicted the contributions to 
the absolute flux and fluctuations of the NIRB by high redshift
($z > 5$) galaxies, some 
of which harboring Pop III stars. This is the most robust and detailed theoretical calculation done so far,
as we simultaneously match 
the LFs and reionization constraints. The simulations include the relevant physics of galaxy formation and a novel treatment of 
chemical feedback, by following the metallicity evolution and implementing the physics of Pop III/Pop II transition based on a critical 
metallicity criterion. It reproduces the observed UV LFs over the redshift range $5 < z < 10$, and extend it to faint 
magnitudes far below the detection limit of current observations.

We directly calculate the stellar emissivity from the simulations.  We use {\tt Starburst99}
to generate metallicity and age dependent 
SED templates, then calculate the luminosity for each galaxy according to its current star formation rate, stellar age and metallicity, 
instead of using a constant metallicity and average main sequence spectrum template. Except for the mass range of the IMF which has 
already been fixed in the simulation, there are no other free parameters in the calculation of the emissivity.

By comparing the number of ionizing photons produced per baryon in collapsed objects,
$f_\star N_\gamma$, in the simulation and the 
ionizing photon rate $N_{\rm ion} \approx f_{\rm esc} f_\star N_\gamma$ deduced from observationally 
constrained reionization models,
we obtained the evolution history of the escape fraction of ionizing photons, $f_{\rm esc}(z)$. We find $f_{\rm esc}\approx 1$ at $z > 11$, 
decreasing to $\approx 0.05$ at $z=5$. This escape fraction is used to renormalize the nebular emission
of Pop III and Pop II stars in the emissivity.

Pop III stars are unlikely to be responsible for the observed NIRB residual, and their contribution 
is very small, making up $<1$\% of the total absolute flux in our calculation.
This is the natural result of the much lower star formation rate of Pop III stars compared 
with Pop II stars in the simulation, since even metals from a single Pop III star could enrich 
above the critical metallicity a large amount
of gas around it \citep{2007MNRAS.382..945T}. The formation of Pop III stars is 
regulated by such a chemical feedback mechanism, which limits their contribution
to the NIRB. However, a rapid Pop III-Pop II transition brings also a little advantage in terms of integrated 
emissivity, due to the longer lifetime of Pop II stars \citep{2012ApJ...756...92C}.

We predict that in the wavelength range $1.0-4.5~\rm \mu m$, the NIRB flux
from $z > 5$ galaxies (and their Pop III stars) 
is $\sim 0.2-0.04~\rm nW/m^2/sr$,
while the fluctuation strength is about $\delta F = 0.01-0.002~\rm nW/m^2/sr$ at $l=2000$.
If we remove galaxies down to $m_{\rm lim} = 28$, the above flux level is only slightly reduced; however, by 
comparing with \cite{2012ApJ...752..113H}, we find that the flux from $z<5$ dramatically decreases and 
the remaining becomes comparable
to the predicted signal of $z > 5$ galaxies.
This implies that in principle it is possible to get the signal from reionization sources 
by subtracting galaxies down to a certain magnitude.

The relative fluctuation amplitude, $\delta F/F$, at  $l=2000$ is $\sim4\%$, almost independent of the wavelength.
This ratio may be helpful to investigate the clustering features of the sources contribute to the 
NIRB, since the intrinsic properties of galaxies 
almost cancel out. Despite the difficulties in measuring the absolute flux accurately, it could be treated as a quality indicator 
in the data reduction process: if a much higher/lower ratio is obtained from the data, this might suggest that a more careful analysis
work is required to extract the genuine contribution from reionization sources.

In spite of being accurate and consistent with the observed LFs 
and reionization data, thus offering a robust prediction of the NIRB
contribution from high-$z$ galaxies which likely reionized the universe, 
a puzzling question remains:
the predicted fluctuations are considerably lower than the observed 
values, indicating that in addition 
to the contribution from the expected high-$z$ galaxy population (and Pop III stars), we should invoke some other -- yet unknown -- 
missing component(s) or foreground(s) which dominates
the currently observed source-subtracted NIRB.
Moreover, the angular clustering of this 
missing component must be very 
similar to that of the high redshift galaxies and extends to degree scales. 
Obviously, this component/foreground must be identified and  
removed before we are ready to exploit the NIRB to study reionization sources. 
 On the other hand, sources located at $5 < z < 8$ 
provide about 90\% of the flux from all sources with $z > 5$ in our simulation; most of them
are the faint galaxies currently undetected
by deep surveys. Thus, if the above mentioned additional spurious sources/foregrounds can be removed reliably, the NIRB 
will become the primary tool to investigate the properties of the reionizing sources.  

\section*{ACKNOWLEDGMENTS}
It is a pleasure to acknowledge intense discussions and data exchange with A. Cooray, 
K. Helgason, E. Komatsu, T. Matsumoto, R. Thompson, S. Kashlinsky, S. Mitra and
T. Choudhury. AF thanks
UT Austin for support and hospitality as a Centennial B. Tinsley Professor
and the stimulating atmosphere of the NIRB Workshop organized by the Texas 
Cosmology Center.
BY and XC also acknowledges the support of the NSFC grant
11073024, the MoST Project 863 grant 2012AA121701, and the Chinese Academy of
Science Knowledge Innovation grant KJCX2-EW-W01.


\appendix
\section{AN ANALYTICAL DERIVATION OF THE EMISSIVITY}\label{appendix}

At redshift $z$, the comoving emissivity of 
stellar population at frequency $\nu$ is given by the following integral
\citep{2006ApJ...646..703F}:
\begin{equation}
\epsilon(\nu,z)=\frac{1}{4\pi}\int_{m_1}^{m_2} L_\nu(m)n_{\star}(m)dm,
\label{eA}
\end{equation}
where $L_{\nu}(m)$ is the specific luminosity of a star with mass $m$, 
$m_1$ is the minimum mass of stars while $m_2$ is the maximum mass,
$n_{\star}(m)$ 
is the number density of {\it shining} stars between $m$ and $m+dm$, which is 
written as
\begin{equation}
n_{\star}(m)=\int_{t(z)-\tau(m)}^{t(z)}\dot{n}_\star(m,t^\prime)dt^\prime,
\label{Ns}
\end{equation}
where $t(z)$ is the age of the universe at redshift $z$, $\tau(m)$ is the lifetime of 
a star with mass $m$.
For Pop II stars with metallicity 1/50~$Z_\odot$,
useful fitting formulae for these quantities as a function of $m$
are collected in \cite{2006ApJ...646..703F}.
Eq. (\ref{Ns}) means that only stars formed between $t(z)-\tau(m)$
and $t(z)$
emit photons at time $t(z)$. The formation rate of stars with mass between $m$ and $m+dm$,
$\dot{n}_{\star}(m,t^\prime)$, is
\begin{equation}
\dot{n}_{\star}(m,t^\prime)=\frac{\dot{\rho}_\star(t^\prime)}{m_\star}f(m),
\end{equation}
in which $f(m)$ is the normalized 
stellar IMF, i.e., $\displaystyle \int_{m_1}^{m_2} f(m)dm = 1$ and $\displaystyle m_\star = \int_{m_1}^{m_2} mf(m)dm$, while 
\begin{equation}
\dot{\rho_\star}(t^\prime)=f_\star\frac{\Omega_b}{\Omega_m}\int^\infty_{M_{\rm min}}M\frac{d^2n}{dMdt^\prime}(M,t^\prime)dM
\label{msfr}
\end{equation}
is the comoving star formation rate density in halos with mass above $M_{\rm min}$,
provided a fraction $f_\star$ of baryons are converted into stars.

Two approximate solutions can be found under particular
circumstances. If the star formation rate is almost constant over the 
time interval $\tau(m)$,
i.e., $\tau(m) < t_{\rm SF}(z)$, where the star formation time scale 
$\displaystyle t_{\rm SF}(z)=\left
  [\frac{\dot{\rho}_\star(z)}{\rho_\star}\right ]^{-1}$, then
we can make ``Approximation 1", i.e.,
$$\int_{t(z)-\tau(m)}^{t(z)}\dot{\rho}_\star(t^\prime)dt^\prime\approx\dot{\rho}_\star[t(z)]\tau(m),$$
and the emissivity is approximated as \citep{2006ApJ...646..703F,2010ApJ...710.1089F}
\begin{equation}
\epsilon(\nu,z)=\frac{1}{4\pi}\frac{\dot{\rho}_\star(z)}{m_\star}\int_{m_1}^{m_2} L_\nu(m)\tau(m)f(m)dm,
\label{a1}
\end{equation}
which is usually used for relative massive stars
with short lifetime.

On the other hand, if $\tau(m)$ is longer than the age of the 
universe (this is true for stars of smaller mass, and means that no stars die),
then we can use ``Approximation 2",
\begin{equation}
\int_{t(z)-\tau(m)}^{t(z)}\frac{\dot{\rho}_\star(t^\prime)}{m_\star}dt^\prime=\int_0^{t(z)}\frac{\dot{\rho}_\star(t^\prime)}{m_\star}dt^\prime=\frac{\rho_\star(z)}{m_\star},
\end{equation}
the emissivity becomes \citep{2006ApJ...646..703F} 
\begin{equation}
\epsilon(\nu,z)=\frac{1}{4\pi}\frac{\rho_\star(z)}{m_\star}\int_{m_1}^{m_2} L_\nu(m)f(m)dm.
\label{a2}
\end{equation}
This also holds true if $\tau(m)$ is much longer than the star
  formation time scale $t_{\rm SF}(z)$, i.e. the death of stars is less significant compared with 
the formation of new stars, so that
\begin{equation}
\int_{t(z)-\tau(m)}^{t(z)}\frac{\dot{\rho}_\star(t^\prime)}{m_\star}dt^\prime=\frac{\Delta \rho_\star(z)}{m_\star}\approx\frac{\rho_\star(z)}{m_\star},   
\end{equation}
and the emissivity could also be approximated as Eq. (\ref{a2}).

In reality, a galaxy is composed of stars with different mass; some of them may have lifetime 
longer than $t_{\rm SF}$, while others not. In this case a ``Hybrid" approximation
could be used,
\begin{align}
\epsilon(\nu,z)=&\frac{1}{4\pi}\left[
\frac{\rho_\star(z)}{m_\star}\int_{m_1}^{m_t} L_\nu(m)f(m)dm+\right.\nonumber \\
&\left.\frac{\dot{\rho}_\star(z)}{m_\star}\int_{m_t}^{m_2}
L_\nu(m)\tau(m)f(m)dm\right], 
\label{ah}
\end{align}
where $m_t$ is the stellar mass determined by the condition $\tau(m_t) = t_{\rm SF}$. 

\begin{figure}
\begin{center}
\includegraphics[scale=0.4]{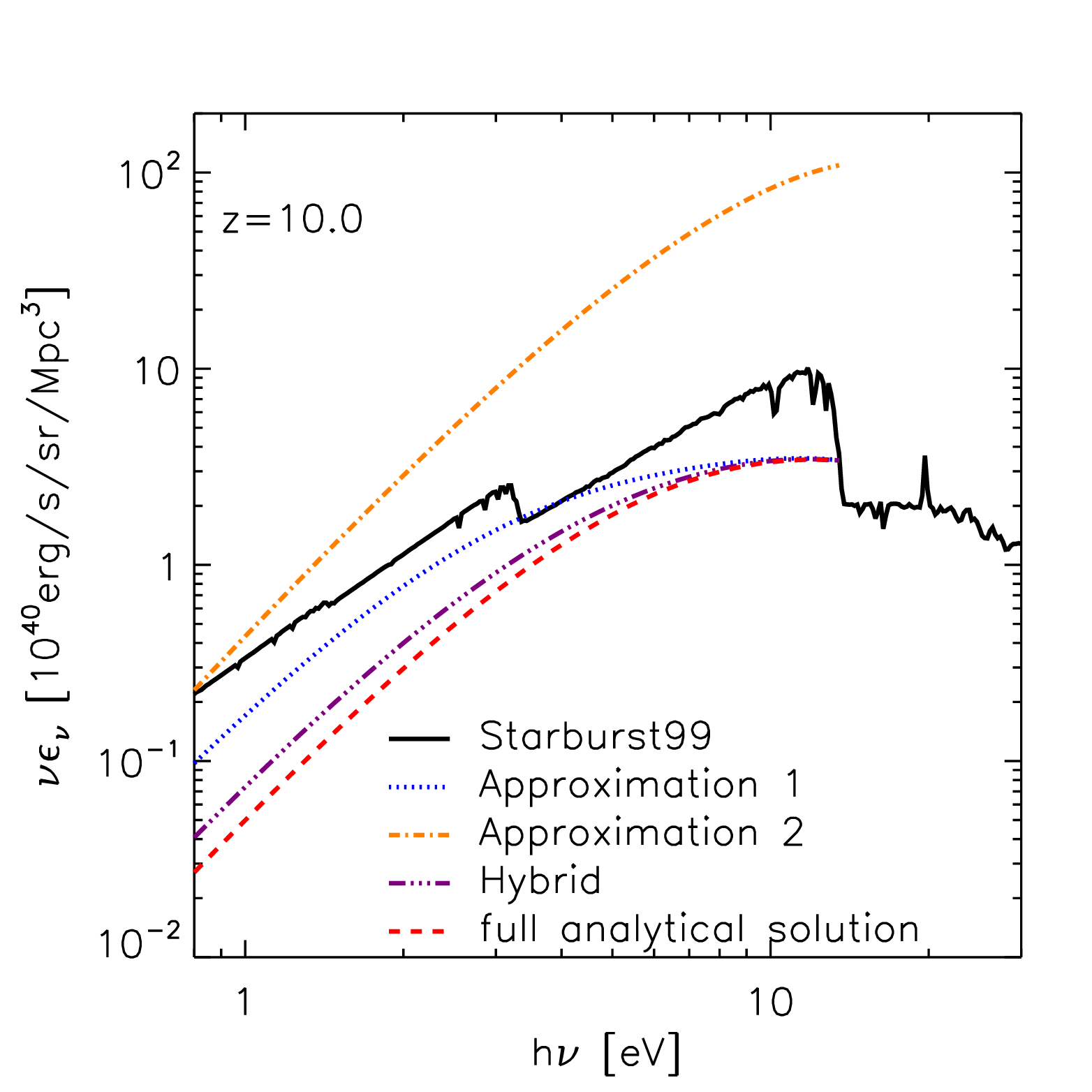}
\caption{Emissivity of Pop II stars at redshift 10.0. The solid line is the result obtained by using the spectrum template 
from {\tt Starburst99} 
used here, the dotted line is for the ``Approximation 1"; dash-dotted line refers to ``Approximation 2". 
The dash-dotted-dotted-dotted line corresponds to the ``Hybrid" approximation; finally the dashed line 
refers to the full analytical 
solution of Eqs (\ref{eA}-\ref{msfr}) using fitting formulae (see text).
}
\label{nuenu}
\end{center}
\end{figure}

We compare the full analytical solution of Eqs. (\ref{eA}-\ref{msfr}) with these three 
approximate solutions. Furthermore, we will also consider the emissivity
obtained by adopting the {\tt Starburst99} template at $Z=1/50~Z_\odot$ 
instead of the simplified fitting
formula given in \cite{2006ApJ...646..703F}. In this case the emissivity is
given by
\begin{equation}
\epsilon(\nu,z)=\frac{1}{4\pi}\int_0^{t(z)}L_{\nu,{\rm SB99}}[z, t(z)-t^\prime]\dot{\rho}_\star(t^\prime)dt^\prime,
\end{equation}
where $L_{\nu,{\rm SB99}}$ is the luminosity per unit mass
(note that here for integration purposes we use the burst star formation model)
from {\tt Starburst99}.

In our work, the mass range of Pop II stars is $0.1-100~M_\odot$,
while the fitted formula of the main sequence age used in
\cite{2006ApJ...646..703F}, \cite{2010ApJ...710.1089F} and 
\cite{2012ApJ...756...92C}
(taken from \citealt{2002A&A...382...28S}) is based on data of massive 
stars. To avoid introducing more uncertainties, in this comparison we 
adopt a mass range $1-100~M_\odot$ for Pop II stars. We checked that for 
Pop II stars with mass 1~$M_\odot$, the fitted main sequence age 
still agrees with \cite{2000A&AS..141..371G}.

Since Pop II stars are found to contribute much more than Pop III stars
to the NIRB (see Figure \ref{Inu0}), and 
stellar emission is the dominant component,
we neglect here the nebular emission.
 $L_\nu$ can then be represented by a blackbody spectrum, and we truncate
it at $h\nu = 13.6$~eV, \citep{2006ApJ...646..703F,2010ApJ...710.1089F,
2012ApJ...756...92C}.
 
We plot the emissivity at $z=10$ calculated by different methods
in Figure \ref{nuenu} for a star formation efficiency 
$f_\star = 0.01$ and a minimum mass $M_{\rm min} = 10^6~M_\odot$.
It is not surprising that ``Approximation 2"
overestimates the emissivity, since it assumes that none of the stars die.
However, we find that for the stellar mass range
$1-100~M_\odot$,
``Approximation 1" also overestimates the emissivity when 
$h\nu < 8$~eV,
because of the contribution of low mass stars whose
  lifetime is even longer than the age of the universe at
that redshift, so that $\tau(m) < t_{\rm SF}$ is not fulfilled.
However, high energy photons  come mainly from massive 
stars, which satisfy $\tau(m) < t_{\rm SF}$. 
So at high energies, ``Approximation 1" results agree 
well with the full analytical solution.
The ``Hybrid" approximation however, is more accurate through
the whole range of energy shown in Figure \ref{nuenu}; yet, the results still
deviate from the full analytical solution.

However, the emissivity calculated from the template of {\tt Starburst99} 
is still higher than the results obtained from the full analytical solution of Eqs. 
(\ref{eA}-\ref{msfr}) with fitting formulae. This is mainly due to the full stellar evolutionary tracks 
used by {\tt Starburst99}, which extend beyond the zero-age main sequence (ZAMS) stage. For example, the 
luminosity of a 7~$M_\odot$ star with metallicity 1/50~$Z_\odot$ at the end of the main sequence
is three times 
as large as the ZAMS luminosity; at the end of its evolution, the luminosity is $\sim10\times$ higher 
compared to  ZAMS luminosity. The analogous value for a 100~$M_\odot$ star of the same metallicity, 
is about $2\times$ the ZAMS luminosity.
\end{document}